\documentstyle[12pt]{article}
\newcommand{\bce}{\begin{center}}
\newcommand{\ece}{\end{center}}
\newcommand{\beq}{\begin{equation}}
\newcommand{\eeq}{\end{equation}}
\newcommand{\bea}{\vspace{0.25cm}\begin{eqnarray}}
\newcommand{\eea}{\end{eqnarray}}

\newcommand{\ba}{\begin{array}}
\newcommand{\ea}{\end{array}}

\newcommand{\r}{\mbox{{\boldmath $\rho$}}}

\newcommand{\doublespace}{
    \renewcommand{\baselinestretch}{1.6}\large\normalsize}

\def\lsim{\mathrel{\rlap{\lower4pt\hbox{\hskip1pt$\sim$}}
    \raise1pt\hbox{$<$}}}	  %less than or approx. symbol
\def\gsim{\mathrel{\rlap{\lower4pt\hbox{\hskip1pt$\sim$}}
    \raise1pt\hbox{$>$}}}	  %greater than or approx. symbol

\def\beq{\begin{equation}}
\def\endeq{\end{equation}}
\def\bea{\begin{eqnarray}}
\def\arr{\begin{eqnarray}}
\def\eea{\end{eqnarray}}

\def\q2{$Q^{2}$}
\def\s2{2$S$}

\begin{document}
\doublespace
\begin{titlepage}
%____________________________________________________________
\vspace*{-2cm}
\begin{flushright}
{\large
%\bf
March 1997\\
LPTHE-Orsay 97-09
}
%\,\,30 October 1996\,\,\,\,\,\,\,\,}
\end{flushright}

\bigskip
  \vspace{1.5 cm}

%------------------------------------------------------------

%\thispagestyle{empty}
\vspace*{-2cm}

\bigskip

\begin{center}

  {\Large \bf
Radiative energy loss of high energy quarks in finite-size
nuclear matter and quark-gluon plasma
}

  \vspace{1.5 cm}

  {\Large    B. G. Zakharov }
  \bigskip
  \bigskip

{\it
Laboratoire de Physique Th\'eorique et Hautes Energies,
B\^atiment 211, Universit\'e de Paris-Sud,\\
91405 Orsay Cedex, France \medskip\\

 L. D. Landau Institute for Theoretical Physics,
	GSP-1, 117940,\\ ul. Kosygina 2, 117334 Moscow, Russia
	\medskip\\}

\vspace{.5cm}

  {\bf\large Abstract}
\end{center}
The induced gluon radiation of a high energy quark
in a finite-size QCD medium is studied.
For a sufficiently energetic quark produced inside a medium
we find the radiative energy loss $\Delta E_{q}\propto L^{2}$,
where $L$ is the distance passed by quark in the
medium. It has a weak dependence on the initial quark energy
$E_{q}$. The $L^{2}$ dependence turns to $L^{1}$ as the quark energy
decreases.
Numerical calculations
are performed for a cold nuclear matter and a hot quark-gluon
plasma.
For a quark incident on a nucleus we predict
$\Delta E_{q}\approx 0.1 E_{q}(L/10\,\mbox{fm})^{\beta}$,
with $\beta$ close to unity.

\vspace*{3cm}

\noindent

\end{titlepage}

\newpage
%\pacs{PACS numbers: 13.60.Le, 12.38.Qk, 25.20.Lj}

%\narrowtext
%\twocolumn
%\doublespace
The radiative energy loss of a high energy parton
in a QCD medium
%has recently attracted much attention
is under active investigation nowadays
%\cite{GW,BD1,BGZ1,BD23}.
[1-5].
In classical electrodynamics the
radiation of a charged particle in a dense medium was first considered
long ago by Landau and Pomeranchuk \cite{LP}. The quantum treatment
of this phenomenon was given by Migdal \cite{Migdal}.
In Ref. \cite{BGZ1} (see also \cite{BGZ2}) we developed a new path
integral approach to the bremsstrahlung in a dense medium applicable
both in QED and QCD.
In the present paper we evaluate within the formalism of Ref.
\cite{BGZ1} the radiative energy
loss of a fast quark, $\Delta E_{q}$, propagating through a
finite-size uniform QCD medium.
We consider both a cold nuclear matter and a hot
quark-gluon plasma (QGP).
Following \cite{GW} we model QGP by a system of static scattering
centres described by
the Debye screened potential $\propto \exp(-r\mu_{D})/r$, where
$\mu_{D}$ is the color screening mass. For the screening
mass we use perturbative formula
$\mu_{D}=(1+n_{F}/6)^{1/2} g_{s}\,T$
\cite{Shuryak1}, where $g_{s}=\sqrt{4\pi \alpha_{s}}$ is the QCD
coupling constant, $T$ is the temperature of QGP.
We assume that a fast quark produced at $z=0$ through a hard
mechanism propagates in a medium of extent $L$ along $z$ axis.

Neglecting the multigluon emission the radiative energy loss can be
written as
\beq
\Delta E_{q}=E_{q}\int\limits_{0}^{1} dx x\frac{dP}{dx}\,,
\label{eq:1}
\eeq
where $E_{q}$ is the initial quark energy, $x$ is the Feynman
variable for the radiated gluon, and $dP/dx$ is the probability
of gluon radiation as function of $x$. In the approach of Ref. \cite{BGZ1}
an evaluation of $dP/dx$  is reduced to solving a
two-dimensional Schr\"odinger
equation in the impact parameter space. The longitudinal
coordinate $z$ plays the role of time. This Schr\"odinger
equation describes evolution of the light-cone wave function of a
spurious three-body $q\bar{q}g$ color singlet system. The relative
positions of the
constituents of the $q\bar{q}g$ system in the impact parameter space
are $\r_{q}=-\r x$, $\r_{\bar{q}}=0$, $\r_{g}=(1-x)\r$.
The corresponding Hamiltonian has the form
\beq
H=\frac{{\mbox{\bf{p}}}^2}{2\mu(x)}+v(\r,z)\,,
\label{eq:2}
\eeq
\beq
v(\r,z)=-i\frac{n(z)\sigma_{3}(\rho,x)}{2}\,.
\label{eq:2p}
\eeq
Here
$
\mu(x)=E_{q}x(1-x)\,
$
is the reduced "Schr\"odinger mass",
$n(z)$ is the medium density, and
$\sigma_{3}(\rho,x)$ is the cross section of interaction
of the $q\bar{q}g$ system with a medium constituent
(color centre for QGP and nucleon for nuclear matter).
In the case of QGP on the rhs of (\ref{eq:2p})
summation over triplet (quark) and octet (gluon)
color states is implicit.

In order to simplify the analysis we
neglect the $q\rightarrow qg$ spin-flip transitions
which give a small contribution to the energy loss. Then
the radiation rate is given by
\cite{BGZ1}
\beq
\frac{d P}{d x}=2\mbox{Re}\!
\int\limits_{0}^{\infty}\! d \xi_{1}\!
\int\limits_{\xi_{1}}^{\infty}d \xi_{2}
\exp\left[-\frac{i(\xi_{2}-\xi_{1})}{L_{f}}\right]
g(\xi_{1},\xi_{2},x)\left[K(0,\xi_{2}|0,\xi_{1})
-K_{v}(0,\xi_{2}|0,\xi_{1})\right]\,.
\label{eq:4}
\eeq
Here the generalization of the QED vertex operator of Ref. \cite{BGZ1}
to QCD reads
\beq
g(\xi_{1},\xi_{2},x)=
\frac{\alpha_{s}[4-4x+2x^{2}]}{3x}\cdot
\frac{{\mbox{\bf p}}(\xi_{2})\cdot{\mbox{\bf p}}(\xi_{1})}{\mu^{2}(x)}
\,,
\label{eq:5}
\eeq
$K$
is the Green's function for the Hamiltonian (\ref{eq:2}), $K_{v}$
is the vacuum Green's function,
$
L_{f}=2E_{q}x(1-x)/[m_{q}^{2}x^{2}+m_{g}^{2}(1-x)]
$
is the so called gluon formation length (time),
$m_{q}$ is the quark mass and $m_{g}$ is the mass of radiated gluon.
The latter plays the role of an infrared
cutoff removing contribution of the long-wave gluon
excitations which cannot be treated perturbatively.
In contrast to
the expression of Ref. \cite{BGZ1} for the bremsstrahlung spectrum
for an electron
incident on a target of Ref. \cite{BGZ1},
in which the  integration
over $\xi_{1}$ starts from $-\infty$, in (\ref{eq:4})
we integrate over $\xi_{1}$ from $\xi_{1}=0$, i.e. from the
point where a fast quark is produced by hard scattering.

The three-body cross section entering the imaginary potential
(\ref{eq:2p})
can be
expressed
in terms of the dipole cross section for color singlet $q\bar{q}$
pair \cite{NZ1}, $\sigma_{2}(\rho)$,
\beq
\sigma_{3}(\rho,x)=\frac{9}{8}[\sigma_{2}(\rho)
+\sigma_{2}((1-x)\rho))]-\frac{1}{8}\sigma_{2}(x\rho)\,.
\label{eq:3}
\eeq
The radiation rate is dominated by the contribution from
$\rho\lsim 1/m_{g}$ \cite{BGZ1}, where
$\sigma_{2}(\rho)=C_{2}(\rho)\rho^{2}$ and $C_{2}(\rho)$
has a smooth (logarithmic) dependence on $\rho$ \cite{DGM,NZ1}.
This allows one to estimate the energy loss replacing
$C_{2}(\rho)$ by
$ C_{2}(1/m_{g})$.
Then
$\sigma_{3}(\rho,x)=C_{3}(x)\rho^{2}$, with
$C_{3}(x)=\{9[1+(1-x)^{2}]-x^{2}\}C_{2}(1/m_{g})/8$,
and the Hamiltonian (\ref{eq:1})
takes the oscillator form with the
frequency
$$
\Omega=\frac{(1-i)}{\sqrt{2}}
\left(\frac{n(z)C_{3}(x)}{\mu(x)}\right)^{1/2}=
\frac{(1-i)}{\sqrt{2}}
\left(\frac{n(z)C_{3}(x)}{E_{q}x(1-x)}\right)^{1/2}\,.
$$

Making use of the oscillator Green's function
after some algebra one can represent the bremsstrahlung
rate (\ref{eq:4}) in the form
\beq
\frac{dP}{dx}=
L n \frac{d\sigma^{BH}_{}}{dx}S_{}(\eta,l)
\,,
\label{eq:8}
\eeq
where
\beq
\frac{d\sigma^{BH}_{}}{dx}=
\frac{4\alpha_{s}C_{3}(x)(4-4x+2x^{2})}{9\pi
x[m_{q}^{2}x^{2}+m_{g}^{2}(1-x)]}\,,
\label{eq:9}
\eeq
is the Bethe-Heitler cross section.
The suppression factor $S(\eta,l)$,
depending on the dimensionless
variables
\beq
\eta=L_{f}|\Omega|=
\frac{[4 n C_{3}(x) E_{q}x(1-x)]^{1/2}}{m_{q}^{2}x^{2}+m_{g}^{2}(1-x)}
\,,
\label{eq:eta}
\eeq
\beq
l=L/L_{f}=\frac{L[m_{q}^{2}x^{2}+m_{g}^{2}(1-x)]}{2E_{q}x(1-x)}\,,
\label{eq:l}
\eeq
is given by
\beq
S_{}(\eta,l)=S_{}^{(1)}(\eta,l)+S_{}^{(2)}(\eta,l)\,,
\label{eq:11}
\eeq
\beq
S_{}^{(1)}(\eta,l)=
\frac{3}{l\eta^{2}}
\mbox{Re}\,
\int\limits_{0}^{l\eta}dy_{1}
\int\limits_{0}^{y_{1}}dy_{2}
\exp\left(-\frac{iy_{2}}{\eta}\right)
\left\{\frac{1}{y_{2}^{2}}-
\left[\frac{\phi}{\sin(\phi y_{2})}\right]^{2}\right\}\,,
\label{eq:12}
\eeq
\bea
S^{(2)}(\eta,l)=
\frac{3}{l\eta^{2}}
\mbox{Re}\,
\int\limits_{0}^{l\eta}dy_{1}
\int\limits_{0}^{\infty}dy_{2}
\exp\left[-\frac{i(y_{1}+y_{2})}{\eta}\right]
\nonumber\\
\times
\left\{\frac{1}{(y_{1}+y_{2})^{2}}-
\left[\frac{\phi}
{\cos(\phi y_{1})\left(\tan(\phi y_{1})+\phi
y_{2}\right)}\right]^{2}\right\}\,,
\label{eq:13}
\eea
with $\phi=\Omega/|\Omega|=\exp(-i\pi/4)$.
The two terms on the rhs of (\ref{eq:11}) correspond in (\ref{eq:4}) to the
contributions from the integration regions
$\xi_{1}<\xi_{2}<L$
and $\xi_{1}<L<\xi_{2}$, respectively.
The variables in (\ref{eq:12}), (\ref{eq:13}) in terms of those in
(\ref{eq:4}) are
$y_{1}=(L-\xi_{1})|\Omega|$, $y_{2}=(\xi_{2}-\xi_{1})|\Omega|$
(in (\ref{eq:12})) and $y_{2}=(\xi_{2}-L)|\Omega|$ (in (\ref{eq:13})).
In arriving at (\ref{eq:13}) we have used
representation of the first Green's function
in the square brackets in (\ref{eq:4}) through convolution
of the oscillator Green's function (for the interval
$(\xi_{1},L)$) and the vacuum one (for the interval $(L,\xi_{2})$).
Notice that the functional form of
our results at $x\ll 1$ differs
from the one obtained in \cite{BD23} within the soft gluon
approximation.

%------------------------------------------------------------------
In a medium it is either $L_{f}$ or $1/|\Omega|$ which sets
the effective medium-modified formation length
$L_{f}^{'}=\mbox{min}(L_{f},1/|\Omega|)$, which is the typical
value of $\xi_{2}-\xi_{1}$ in (\ref{eq:4}) for $L\gg L_{f}^{'}$.
The finite-size effects come into play only at
$L\lsim L_{f}^{'}$, i.e. $l\lsim l_{0}=\mbox{min}(1,1/\eta)$.
>From (\ref{eq:11})-(\ref{eq:13}) we find
$S(\eta,l)\approx- l^{2}\log l$ as $l\rightarrow 0$.
The source of this suppression of radiation at small $L$
is obvious: the energetic quark produced through a hard mechanism loses
soft component of its gluon cloud and radiation at distances shorter
than the
time required for  regeneration of the quark
gluon field turns out to be suppressed.
For $l\gg l_{0}$ $S(\eta,l)$ reduces to that
for the infinite medium, for which
$S(\eta,l=\infty)\approx 3/\eta\sqrt{2}$ ($\,\eta \gg 1$)
and
$S(\eta,l=\infty)\approx 1-16\eta^{4}/21$ ($\,\eta \ll 1$)
were derived in \cite{BGZ1}. Notice, that according to
(\ref{eq:eta}), (\ref{eq:l})
$\eta \rightarrow 0$
and $l\rightarrow \infty$ as $x\rightarrow 0,1$ and the Bethe-Heitler
regime takes place in these limits.

Before presenting the numerical result, let us consider
the energy loss at a qualitative level.
We begin with the case of a sufficiently large $E_{q}$
such that the maximum value of
$L_{f}^{'}$, $L_{f}^{'}(max)$, is much bigger than $L$.
Taking into account  the finite-size suppression of
radiation at $L_{f}^{'}\gsim L$, we find that $\Delta E_{q}$ is dominated
by the contribution from two narrow regions of $x$:
$x\lsim \delta_{g}\approx L m_{g}^{2}/2l_{0}E_{q}$
and $1-x\lsim \delta_{q}\approx L m_{q}^{2}/2l_{0}E_{q}$.
In both the regions the finite-size effects are marginal
and the energy loss can be estimated using the infinite medium
suppression factor. For instance,
\beq
\Delta E_{q}(x\lsim \delta_{g})\sim
\frac{16\alpha_{s}C_{3}(0)E_{q}Ln}{9\pi m_{g}^{2}}
\int\limits_{0}^{\delta_{g}}dxS(\eta(x),l=\infty)\,.
\label{eq:estimate}
\eeq
Using (\ref{eq:eta}) one can show that $\eta(x\lsim \delta_{g})\lsim 1$
at $L\lsim m_{g}^{2}/2nC_{3}(0)$. In this region of
$L$ in (\ref{eq:estimate}) we can put $S(\eta(x),l=\infty)\approx 1$
and find $\Delta E_{q}\sim 0.25 \alpha_{s}C_{3}(0)n L^{2}$,
which does not depend on the quark energy.
At $L\gg m_{g}^{2}/2nC_{3}(0)$ the typical values of $\eta$
in (\ref{eq:estimate}) are much bigger than unity, and using
the asymptotic formula
for the suppression factor we obtain
$\Delta E_{q}\sim  \alpha_{s}C_{3}(0)n L^{2}$.
Similar analysis for $x$ close to unity
gives the contribution to $\Delta E_{q}$ suppressed by
the factor $\sim 1/4$ as compared to that for small $x$.
Notice that in this $L^{2}$ regime, despite
the $1/m_{g,q}^{2}$ infrared divergence of the Bethe-Heitler
cross section, $\Delta E_{q}$ has only a smooth
$m_{g}$-dependence originating from the factor $C_{3}$.
We emphasize that the above analysis of the origin of the
leading contributions makes it evident
that $L^{2}$ dependence
of $\Delta E_{q}$ cannot be regarded as a consequence of
the Landau-Pomeranchuk-Migdal suppression of the radiation rate
due to small angle multiple scatterings.

The finite-size effects can be neglected
and $\Delta E_{q}$ becomes proportional to $L$
if
%in the whole range of $x$ $L_{f}^{'}\ll L$.
$L_{f}^{'}(max)\ll L$.
If in addition the typical values of $\eta$ are much bigger than unity,
from (\ref{eq:1}), (\ref{eq:8}),
(\ref{eq:9}) along with the
asymptotic form of $S(\eta,l=\infty)$ at $\eta\gg 1$ one can obtain
the following infrared stable result
$\Delta E_{q}\approx 1.1 \alpha_{s}L\sqrt{nC_{3}(0) E_{q}}$.
%

%-----------------------------------------------------
In numerical calculations we take $m_{g}=0.75$ GeV. This value
of $m_{g}$ was obtained in  \cite{BFKL1} from the analysis
of HERA data on structure function $F_{2}$ within
the dipole approach \cite{BFKL2} to the BFKL equation. It is also
consistent
with the nonperturbative estimate \cite{Shuryak2} of the gluon
correlation radius in QCD vacuum.
For scattering of the $q\bar{q}g$ system on a nucleon, we find
from the double gluon model \cite{DGM} $C_{2}(1/m_{g})\sim 1.3-4$ where
the lower and upper bounds correspond to the
$t$-channel
gluon propagators with mass 0.75 and 0.2 GeV, respectively.
The latter choice allows one to
reproduce the dipole
cross section extracted from the data on vector meson
electroproduction \cite{NNPZ}. However, there is every indication
\cite{BFKL1,BFKL2}
that a considerable part of the dipole cross section obtained
in \cite{NNPZ} comes from the nonperturbative effects for which
our approach is not justified. For this reason we take
 $C_{2}(1/m_{g})=2$ which seems to be
plausible estimate for the perturbative component of the
dipole cross section \cite{BFKL1}.
For scattering of the $q\bar{q}g$ system on color
centre
we estimated $C_{2}(1/m_{g})$ using the double gluon formula with the
Debye screened gluon exchanges.
For $T=250$ MeV we obtained
$C_{2}(1/m_{g})\approx 0.5$ for
triplet centre. For octet centre the
result is $C_{A}/C_{F}=9/4$ times larger, here
$C_{A}(C_{F})$ is the octet(triplet) second-order Casimir invariant.
For quark mass, controlling the transverse size of the $q\bar{q}g$
system at $x\approx 1$, we take $m_{q}=0.2$ GeV. Notice that
our prediction for $\Delta E_{q}$ is insensitive to the value of $m_{q}$.

For nuclear matter taking
$n=0.15$ fm$^{-3}$  and $\alpha_{s}=1/2$ for $L\lsim 5$ fm
we obtained $\Delta E_{q}\approx a (L/5\,\mbox{fm})^{\beta}$,
with $a\approx$ 0.55, 1, 1.23 GeV  and $\beta\approx$ 1.5, 1.85, 1.95
for $E_{q}=$10, 50, 250 GeV.
Calculations with $\alpha_{s}=1/3$ for
QGP at $T=250$ MeV yield for the same energies:
$a\approx$ 4.2, 10.2, 14.8 GeV  and $\beta\approx$ 1.2, 1.65, 1.9.
The above values of $\beta$ were determined for $L\lsim 5$ fm.
In the region $5\lsim L \lsim 10$ fm $\beta$ is by 10-20 \% smaller.
At $E_{q}\gsim 250$ GeV $a$ and $\beta$ flatten.
Notice that
$L_{f}^{'}(max)\sim 5-10$ fm for
$E_{q}\sim 10-40$ GeV in nuclear matter, and for $E_{q}\sim 150-600$ GeV
in QGP.
Thus our numerical results say that the onset of $L^{2}$ regime
takes place when $L_{f}^{'}(max)/L\gsim 2$.
The closeness of $\beta$ to unity at $E_{q}=10$ GeV for QGP
agrees with a small value of $L_{f}^{'}(max)$ ($\sim 1$ fm).
We checked that variation of $m_{q}$ gives a
small effect. The $m_{g}$-dependence of $\Delta E_{q}$ becomes weak
at $E_{q}\gsim 50$ GeV. However, it is sizeable for
 $E_{q}\sim 10-20$ GeV.
For instance,
$\Delta E_{q}(m_{g}=0.375)/\Delta E_{q}(m_{g}=0.75)\sim 1.5$
at $E_{q}=10$ GeV, $L\sim 5$ fm.
Our predictions for $\Delta E_{q}$ must be regarded as
rough estimates with uncertainties of at least a factor of 2 in either
direction.
Nonetheless rather large values of $\Delta E_{q}$ obtained
for QGP indicate that the jet quenching
may be an important  potential probe for
formation of the deconfinement phase in $AA$ collisions.

We also studied the energy loss of a fast quark incident on a target.
In this case the radiation by initial quark is allowed and
the lower limit of integration over $\xi_{1}$ in
(\ref{eq:4}) must be replaced by $-\infty$. For the bremsstrahlung
in QED this situation was discussed in \cite{BGZ2}. It was shown
that after expanding the medium Green's function in a series in
the potential the spectrum can be represented as a sum
of the Bethe-Heitler term and an absorptive correction.
For our choice of the gluon mass the absorptive correction
is relatively small.
This means that
$\Delta E_{q}\propto  E_{q} L n \alpha_{s}C_{3}(0)/m_{g}^{2}$.
For nuclear matter in the region $L\lsim 10$ fm the numerical
calculations give
$\Delta E_{q}\approx 0.1 E_{q}(L/10\,\mbox{fm})^{\beta}$
with $\beta\approx 0.9-1$ for $E_{q}\lsim 50$ GeV and
$\beta\approx 0.85-0.9$ for $E_{q}\gsim 200$ GeV.
This result differs drastically from prediction
of Ref. \cite{Brodsky} $\Delta E_{q}\approx 0.25 (L/1\, \mbox{fm})$ GeV.
Our estimate is in a qualitative
agreement with the longitudinal energy flow measured
in hard $pA$ collisions with dijet final state \cite{E609}
and the energy loss obtained from the analysis of the
inclusive hadron spectra in $hA$ interactions \cite{QK}.

I would like to thank
D.~Schiff for discussions and hospitality
at LPTHE, Orsay, where this work was completed.

\end{document}